\documentclass[12pt]{article}

\usepackage{amsmath}  
\usepackage{amssymb}  
\usepackage{graphicx} 
\usepackage{hyperref} 
\usepackage{algorithm} 
\usepackage{algorithmic} 
\usepackage{geometry} 
\usepackage{caption}  
\usepackage{subcaption} 
\usepackage{natbib}  

\title{Game Theory and Multi-Agent Reinforcement Learning : From Nash Equilibria to Evolutionary Dynamics}
\author{
    Neil De La Fuente \\
    Computer Vision Center,UAB \\
    \and
    Miquel Noguer i Alonso \\
  AIFI \\
    \and
    Guim Casadellà \\
    AllRead
}

\date{\today}

\begin{document}

\maketitle

\begin{abstract}
This paper explores advanced topics in complex multi-agent systems building upon our previous work. We examine four fundamental challenges in Multi-Agent Reinforcement Learning (MARL): non-stationarity, partial observability, scalability with large agent populations, and decentralized learning. The paper provides mathematical formulations and analysis of recent algorithmic advancements designed to address these challenges, with a particular focus on their integration with game-theoretic concepts. We investigate how Nash equilibria, evolutionary game theory, correlated equilibrium, and adversarial dynamics can be effectively incorporated into MARL algorithms to improve learning outcomes. Through this comprehensive analysis, we demonstrate how the synthesis of game theory and MARL can enhance the robustness and effectiveness of multi-agent systems in complex, dynamic environments.
\end{abstract}

\section{Introduction}

Multi-Agent Reinforcement Learning (MARL) has emerged as a vital field in Artificial Intelligence (AI), focusing on how multiple agents learn and interact within shared environments. The integration of game theory with MARL provides a robust mathematical framework for understanding strategic behaviors among rational agents. Building on our previous work, "Game Theory and Multi-Agent Reinforcement Learning: A Mathematical Overview," \cite{Noguer2024} we explore advanced topics and address significant challenges inherent in MARL.

We discuss the main challenges MARL faces: nonstationarity, where simultaneous learning and adaptation by agents change the environment's dynamics from any single agent's perspective; partial observability, requiring agents to make decisions under uncertainty due to limited or noisy information; scalability issues from the exponential growth of the joint action space with additional agents, leading to computational and coordination difficulties; and decentralized learning approaches, which aim to mitigate scalability issues by relying on local observations but introduce challenges in ensuring effective coordination without centralized control.

We analyze recent advances in MARL algorithms that have incorporated game-theoretic concepts. Extensions of Deep Q-Networks (DQN) for multi-agent settings and policy gradient methods like Multi-Agent Deep Deterministic Policy Gradient (MADDPG) have been developed to handle both cooperative and competitive interactions. Other approaches such as hierarchical reinforcement learning break down complex tasks into manageable subtasks, improving learning efficiency and scalability in multi-agent systems.

This work provides a comprehensive overview of these advanced topics, emphasizing their mathematical foundations and practical implications. By integrating game-theoretic principles with MARL, we can develop more robust and efficient multi-agent systems capable of operating in complex, dynamic environments. The ongoing research in addressing non-stationarity, partial observability, scalability, and decentralized coordination is paving the way for significant advancements in fields such as finance, robotics and beyond.


\section{Challenges in Multi-Agent Reinforcement Learning}
Multi-Agent Reinforcement Learning (MARL) extends traditional single-agent RL to environments with multiple interacting agents. While this extension offers richer modeling capabilities and potential for more complex and realistic applications, it introduces several significant challenges. This section explores the primary challenges in MARL, including Non-Stationarity, Partial Observability, Scalability with Large Agent Populations, and Decentralized Learning and Coordination. These and some other relevant challenges were presented by \cite{9043893}. 

\subsection{Non-Stationarity}
Non-stationarity in MARL arises because the environment's dynamics are influenced by the evolving policies of multiple agents. Unlike single-agent environments where the transition dynamics remain constant, in MARL, each agent's policy updates can alter the environment and transition dynamics, creating a moving target for other agents.

\subsubsection{Formal Definition}

Formally, consider a stochastic game defined by the tuple \( \langle S, A_1, \dots, A_n, P, R_1, \dots, R_n, \gamma \rangle \), where:
\begin{align*}
    S & \text{ is the set of states,} \\
    A_i & \text{ is the action set for agent } i, \\
    P(s' | s, a_1, a_2, \dots, a_n) & \text{ is the state transition function,} \\
    R_i(s, a_1, a_2, \dots, a_n) & \text{ is the reward function for agent } i, \\
    \gamma & \text{ is the discount factor.}
\end{align*}
Each agent \( i \) follows a policy \( \pi_i(a_i | o_i) \), where \( o_i \) is the observation available to agent \( i \). The joint policy \( \pi = (\pi_1, \pi_2, \dots, \pi_n) \) induces non-stationarity because the environment perceived by any single agent changes as other agents update their policies.

\subsubsection{Challenges and Implications on Learning Algorithms}

The primary challenge posed by non-stationarity is the \textbf{moving target problem}. In MARL, as agents iteratively update their policies \( \pi_i \), the environment dynamics \( P \) and reward functions \( R_i \) shift accordingly. This continual adaptation undermines the convergence guarantees provided by traditional single-agent RL algorithms, which rely on a stationary environment. Mathematically, the optimal policy for an agent is defined as:
\begin{equation}
\pi_i^* = \arg\max_{\pi_i} \mathbb{E}\left[ \sum_{t=0}^{\infty} \gamma^t R_i(s_t, a_{1,t}, a_{2,t}, \dots, a_{n,t}) \mid \pi \right].  
\label{optimal_policy}
\end{equation}

As \( \pi_j \) for \( j \neq i \) evolves, the optimal \( \pi_i^* \) for agent \( i \) shifts accordingly, making it difficult for agents to stabilize their learning processes.

This dynamic interplay leads to several critical implications for learning algorithms:

\paragraph{Convergence and Stability.}
Traditional RL algorithms assume fixed transition dynamics and reward structures, ensuring that policy updates move towards optimality. However, in a non-stationary environment, these assumptions are violated. The Bellman optimality equation, which is the underlying equation in many RL algorithms, becomes time-dependent:
\begin{equation}
Q_i^{\pi}(s, a_i) = \mathbb{E}_{a_{-i} \sim \pi_{-i}, s' \sim P(\cdot|s, a_i, a_{-i})} \left[ R_i(s, a_i, a_{-i}) + \gamma \max_{a'_i} Q_i^{\pi}(s', a'_i) \right].
\label{bellman_optimality}
\end{equation}

Here, \( Q_i^{\pi}(s, a_i) \) depends on the policies \( \pi_{-i} \) of other agents, which are continuously evolving. This interdependence may cause policy oscillations, where agents perpetually adjust their strategies in response to one another without ever stabilizing, difficulting convergence to stable solutions.

\paragraph{Sample Efficiency and Exploration.}
Non-stationarity increases the sample complexity of learning algorithms. In a stationary environment, each interaction provides consistent and reliable information for policy updates. In contrast, in a non-stationary setting, the relevance of each sample diminishes as policies evolve, requiring agents to gather more and more interactions to obtain sufficiently accurate and up-to-date information about the environment. This increased data hunger turns the learning process slow and requires more robust exploration strategies to ensure that agents can adapt effectively to the shifting dynamics.

\paragraph{Credit Assignment and Policy Evaluation.}
Determining the individual contribution of an agent's actions to the received rewards becomes more complex in a multi-agent setting. Since rewards are influenced by the collective actions of all agents, accurately attributing rewards to specific actions is challenging. This ambiguity complicates the gradient estimation required for policy updates. For example, in policy gradient methods, the gradient for agent \( i \) is given by:

\begin{equation}
\nabla_{\theta_i} J(\pi_i) = \mathbb{E}\left[ \nabla_{\theta_i} \log \pi_i(a_i \mid o_i) Q_i^{\pi}(s, a_i) \right].
\label{eq:policy_gradient}
\end{equation}

Here, as shown in Equation~\ref{bellman_optimality}, \( Q_i^{\pi}(s, a_i) \) implicitly depends on the policies \( \pi_{-i} \) of other agents, making it difficult to isolate the effect of agent \( i \)'s actions on its own rewards. This interdependence makes it difficult for agents to accurately update their policies, as they must separate their actions' contributions from those of others.

\paragraph{}
Other challenges include the need for adaptive learning rates and stability mechanisms to manage the variability introduced by evolving policies, as well as ensuring robustness to dynamic interactions and balancing individual objectives with collective goals. These additional complexities burden the capabilities of traditional learning algorithms, showcasing the clear necessity for more sophisticated approaches tailored to the non-stationary nature of MARL environments.

In summary, non-stationarity in MARL poses significant challenges to the convergence, stability, and efficiency of learning algorithms. Addressing these implications requires the development of sophisticated techniques that can account for the dynamic interplay between multiple learning agents, ensuring that policies can adapt effectively while maintaining stability and coherence within the multi-agent system.

\subsection{Partial Observability}
Partial observability occurs when agents do not have access to the complete state of the environment, relying instead on limited or noisy observations \cite{oliehoek}. This limitation necessitates the development of algorithms that can infer hidden state information or make robust decisions under uncertainty.

\subsubsection{Formal Definition}

A Partially Observable Stochastic Game (POSG) extends the Markov game framework by introducing an observation function \( O_i: S \rightarrow \mathcal{O}_i \) for each agent \( i \), where \( \mathcal{O}_i \) is the set of possible observations. At each time step, agent \( i \) receives an observation \( o_i \in \mathcal{O}_i \) that provides partial information about the true state \( s \in S \):
\begin{equation}
o_i = O_i(s).
\label{eq:partial_observation}
\end{equation}
This partial observability means that agents must make decisions based on incomplete information, which can obscure the true state and the intentions of other agents.

\subsubsection{Challenges and Implications on Learning Algorithms}

Partial observability introduces significant complexities in the decision-making and learning processes of agents within a multi-agent system. The inherent uncertainty and incomplete information require agents to infer the underlying state of the environment, which has deep implications for the design and effectiveness of learning algorithms in MARL.

\paragraph{State Inference and Belief States.}
In a partially observable environment, agents must maintain and update a belief state \( b_i(s) = P(s | \tau_i) \), where \( \tau_i \) represents the history of observations and actions for agent \( i \). This belief state encapsulates the agent's estimation of the true state based on past interactions:
\begin{equation}
    b_i(s) = P(s | o_{i,1}, a_{i,1}, a_{i,2}, \dots, o_{i,t}, a_{i,t}).
\label{eq:belief_state}
\end{equation}
Maintaining accurate belief states is computationally demanding, as it requires integrating information over time and handling the probabilistic nature of state estimations. Errors in belief state estimation can lead to suboptimal decision-making, as agents base their actions on incorrect or incomplete state information.

\paragraph{Non-Markovian Dependencies.}
Partial observability often introduces non-Markovian dependencies, where the optimal policy depends not only on the current observation but also on the history of past observations and actions. Mathematically, the policy \( \pi_i(a_i | \tau_i) \) must consider the entire trajectory \( \tau_i \) to make informed decisions:
\begin{equation}
\pi_i^*(a_i | \tau_i) = \arg\max_{\pi_i} \mathbb{E}\left[ \sum_{t=0}^{\infty} \gamma^t R_i(s_t, a_{1,t}, a_{2,t}, \dots, a_{n,t}) \mid \pi_i, \pi_{-i}, \tau_i \right].
\label{eq:optimal_policy_partial}
\end{equation}
This dependency complicates the learning process, as agents must effectively summarize and utilize historical information to maintain a coherent representation of the environment's state.

\paragraph{Increased Uncertainty and Exploration Challenges.}
Limited and noisy observations amplify uncertainty in state estimation, making it difficult for agents to predict other agents' actions and the resulting state transitions accurately. This uncertainty complicates both the exploration and exploitation phases of learning. Efficient exploration strategies must balance the need to gather information about the hidden state with the objective of maximizing rewards. Traditional (old-fashioned) exploration methods, such as \(\epsilon\)-greedy or Boltzmann exploration, may be insufficient in partially observable settings, necessitating more sophisticated approaches that account for the uncertainty and information gain.

\paragraph{Memory and Computational Overhead.}
To handle partial observability, learning algorithms often incorporate mechanisms to retain and process historical information, such as recurrent neural networks (RNNs) or Long Short-Term Memory (LSTM) networks. These architectures enable agents to maintain an internal state that captures relevant aspects of past observations and actions. However, this requirement increases both memory usage and computational complexity, as agents must manage and process extensive temporal information to make informed decisions.

\paragraph{Credit Assignment and Policy Evaluation.}
In partially observable environments, accurately attributing rewards to specific actions becomes more challenging due to the lack of complete state information. The credit assignment problem is exacerbated, as agents must discern the impact of their actions on rewards without direct visibility into the underlying state transitions. This ambiguity complicates the estimation of value functions and policy gradients, potentially leading to slower or less accurate policy updates:
\begin{equation}
\nabla_{\theta_i} J(\pi_i) = \mathbb{E}\left[ \nabla_{\theta_i} \log \pi_i(a_i \mid o_i) Q_i^{\pi}(s, a_i) \right].
\label{eq:policy_gradient_partial}
\end{equation}
Here, \( Q_i^{\pi}(s, a_i) \) is influenced by the hidden states and the policies of other agents, making it difficult to isolate the effect of agent \( i \)'s actions on its own rewards.
\paragraph{}
Other challenges include the complexities of policy representation and the potential for convergence and stability issues due to the reliance on partial historical information and potentially biased belief states. These factors make it more demanding for algorithms to effectively learn and optimize policies in partially observable multi-agent environments.

In summary, partial observability in MARL presents multiple challenges that significantly impact the design and effectiveness of learning algorithms. The need for accurate state inference, handling non-Markovian dependencies, managing increased uncertainty, and addressing computational and memory overheads complicates the learning process. Additionally, the difficulties in credit assignment, policy representation, convergence, and robustness to noise further exacerbate these challenges. Addressing these implications requires the development of sophisticated algorithms capable of efficiently summarizing historical information, managing uncertainty, and maintaining stable and effective policy updates in the face of incomplete and noisy observations.

\subsection{Scalability with Large Agent Populations}

Scalability in MARL refers to the ability of learning algorithms to handle environments with a large number of agents. As the number of agents increases, the complexity of the joint action space grows exponentially, posing significant computational and memory challenges \cite{10550936}.

\subsubsection{Formal Definition}

In a multi-agent system with \( n \) agents, each with an action set \( A_i \), the joint action space is defined as:
\begin{equation}
A = A_1 \times A_2 \times \dots \times A_n.
\label{eq:joint_action_space}
\end{equation}
The size of the joint action space is given by:
\begin{equation}
|A| = \prod_{i=1}^{n} |A_i|,
\label{eq:joint_action_size}
\end{equation}
which grows exponentially with the number of agents \( n \). This exponential growth makes it computationally infeasible to evaluate and update policies for all possible action combinations as the number of agents scales.

\subsubsection{Challenges and Implications on Learning Algorithms}

The exponential growth of the joint action space with the number of agents results in substantial computational complexity and memory overhead. As the number of agents \( n \) increases, the joint action space \( A \) defined in Equation~\ref{eq:joint_action_space} becomes exponentially larger, as shown in Equation~\ref{eq:joint_action_size}. This rapid expansion leads to several critical challenges and implications for learning algorithms in MARL:

\paragraph{Computational Complexity and Resource Demands}
Evaluating and updating policies across an exponentially growing joint action space becomes increasingly resource-intensive. The computational burden comes from the need to process a vast number of possible action combinations, making real-time decision-making and policy updates impractical if resources are constrained. Mathematically, the Bellman equation for the joint action-value function \( Q(s, \mathbf{a}) \) is expressed as:
\begin{equation}
Q(s, \mathbf{a}) = R(s, \mathbf{a}) + \gamma \sum_{s'} P(s' | s, \mathbf{a}) \max_{\mathbf{a}'} Q(s', \mathbf{a}').
\label{eq:bellman_joint_action}
\end{equation}
Here, \( \mathbf{a} = (a_1, a_2, \dots, a_n) \) represents the joint action of all agents. The summation over all possible joint actions \( \mathbf{a}' \) in the Bellman equation becomes intractable as \( n \) grows, hindering efficient policy evaluation and updates.

\paragraph{Memory Constraints}
Storing value functions or policies over large joint action spaces demands substantial memory resources. Each agent's policy must account for the large number of combinations of actions taken by other agents, leading to high-dimensional representations that are often impractical to store and manage. This memory bottleneck limits the scalability of MARL algorithms, particularly in environments where computational and memory resources are constrained.

\paragraph{Sample Inefficiency and Exploration Challenges}
With an exponentially larger joint action space, the sample complexity of learning algorithms escalates. Agents require a significantly higher number of interactions to adequately explore and learn effective policies within the vast state-action space. This increased demand for samples slows down the learning process and makes it more resource-intensive. Moreover, efficient exploration becomes way more challenging, as the likelihood of encountering informative experiences decreases with the growth of the joint action space.

\paragraph{Optimization Difficulties}
High-dimensional action spaces complicate gradient-based optimization methods, potentially leading to slower convergence rates or convergence to suboptimal policies. The increased complexity makes it harder for optimization algorithms to navigate the policy landscape effectively, as the gradients may become noisy or less informative in the presence of numerous interacting agents.

\paragraph{Inter-Agent Dependencies and Coordination Overhead}
As the number of agents increases, the dependencies between their actions become more complicated. Coordinating actions and maintaining consistent group behavior require sophisticated mechanisms to manage these dependencies, increasing the computational and algorithmic overhead. Ensuring that policies remain coherent and cooperative across a large number of agents is inherently complex, promoting scalability issues.

\paragraph{Inference Latency and Real-Time Decision Making}
The time required to infer optimal actions from a vast joint action space can lead to significant latency, making real-time decision-making infeasible. In time-sensitive applications, such as autonomous driving or real-time strategy games, this latency can severely impact the performance and usability of MARL systems.
\paragraph{}
Other challenges include the difficulty in ensuring policy generalization across different agent configurations and the increased communication overhead in scenarios where inter-agent communication is necessary. These factors collectively strain the capabilities of traditional MARL algorithms, showcasing the need for more scalable and efficient approaches.

In summary, scalability with large agent populations in MARL presents many challenges that significantly impact the computational efficiency, memory requirements, and overall effectiveness of learning algorithms. The exponential growth of the joint action space leads to increased computational complexity, memory constraints, sample inefficiency, optimization difficulties, intricate inter-agent dependencies, and inference latency. Addressing these challenges is essential for the development of robust and scalable multi-agent systems capable of operating effectively in environments with a large number of interacting agents.
   
\subsection{Decentralized Learning and Coordination}

Decentralized learning involves that each agent independently learns and updates its policy based solely on local observations and interactions, without reliance on a central coordinator \cite{oroojlooyjadid2021reviewcooperativemultiagentdeep}. Although this approach improves scalability and robustness, it introduces significant challenges in coordination and consistency among agents.

\subsubsection{Formal Definition}

In decentralized MARL, each agent \( i \) maintains its own policy \( \pi_i(a_i | o_i) \) and updates it based on its local observations and experiences. There is no central authority or shared memory that provides global state information or coordinates policy updates. Formally, each agent seeks to optimize its own objective:
\begin{equation}
\pi_i^* = \arg\max_{\pi_i} \mathbb{E}\left[ \sum_{t=0}^{\infty} \gamma^t R_i(s_t, a_{1,t}, a_{2,t}, \dots, a_{n,t}) \mid \pi_i, \pi_{-i} \right],
\label{eq:decentralized_objective}
\end{equation}
where \( \pi_{-i} \) denotes the policies of all other agents.

\subsubsection{Challenges and Implications on Learning Algorithms}

Decentralized learning introduces a lot of challenges that significantly impact the design and effectiveness of learning algorithms in MARL. The lack of centralized coordination forces agents to operate based on limited local information, leading to complexities in coordination, credit assignment, and policy consistency. These challenges have deep implications on how learning algorithms must be structured and operate in decentralized settings.

\paragraph{Missing Global Information and Coordination.}
Without access to global state information or synchronized policy updates, agents must rely solely on their local observations to make decisions. This limitation hampers the ability to coordinate effectively, as agents cannot anticipate the actions of others with certainty. Mathematically, the joint policy \( \pi = (\pi_1, \pi_2, \dots, \pi_n) \) lacks a centralized framework to ensure coherent policy updates as shown in Equation~\ref{eq:decentralized_objective}.
This decentralized approach leads to independent policy optimizations that may not align with the collective objectives of the multi-agent system.

\paragraph{Credit Assignment Problem.}
The credit assignment problem becomes more pronounced in decentralized settings. Since rewards are influenced by the collective actions of all agents, it becomes difficult to determine the individual contribution of each agent's actions to the received rewards. This ambiguity complicates the gradient estimation required for policy updates. For example, in policy gradient methods, the gradient for agent \( i \) is given by:
\begin{equation}
\nabla_{\theta_i} J(\pi_i) = \mathbb{E}\left[ \nabla_{\theta_i} \log \pi_i(a_i \mid o_i) Q_i^{\pi}(s, a_i) \right].
\label{eq:decentralized_policy_gradient}
\end{equation}
Here, \( Q_i^{\pi}(s, a_i) \) implicitly depends on the policies \( \pi_{-i} \) of other agents, making it challenging to isolate the effect of agent \( i \)'s actions on its own rewards. This interdependence impedes accurate policy updates, as agents must decouple their actions' contributions from those of others without centralized oversight.

\paragraph{Policy Consistency and Stability.}
Ensuring policy consistency across agents is a significant challenge in decentralized learning. Each agent's policy update can alter the environment dynamics perceived by others, leading to a continuous interplay that complicates the stabilization of policies across the system. This interdependence can result in conflicting strategies or suboptimal collective behavior, reducing the overall effectiveness of the multi-agent system. Mathematically, the update rule for one agent affects the learning signals of others:

\begin{equation}
\pi_j^{(t+1)} = \arg\max_{\pi_j} \mathbb{E}\left[ \sum_{t=0}^{\infty} \gamma^t R_j(s_t, a_{1,t}, a_{2,t}, \dots, a_{n,t}) \mid \pi_j, \pi_i^{(t)} \right].
\label{eq:policy consistency}
\end{equation}
Such mutual dependencies can lead to oscillations or divergence in policy updates, preventing convergence to stable and optimal policies for each of the agents, both individually and collectively.

\paragraph{Exploration-Exploitation Trade-Off.}
In decentralized settings, managing the trade-off between exploration and exploitation becomes more complex. Agents must explore the state-action space efficiently while exploiting known good strategies, all based on limited, potentially noisy and certainly biased local observations. Traditional exploration strategies become insufficient, as the decentralized nature limits the ability to coordinate exploration efforts. Efficient exploration requires agents to balance the need for information gathering with the pursuit of rewards, often without a centralized mechanism to guide this balance.

\paragraph{Increased Learning Complexity.}
Decentralized learning algorithms often require more sophisticated policy representations to handle the uncertainty and variability introduced by the independent learning processes of other agents. Policies may need to incorporate mechanisms for adapting to changing environment dynamics and the evolving behaviors of other agents, increasing the complexity of the learning architecture. This added complexity can lead to longer training times and greater computational resource demands, making the learning process more challenging and less efficient.
\paragraph{}
Other challenges include maintaining robustness to dynamic interactions and ensuring that individual policy updates do not negatively impact the collective behavior of the multi-agent system. These factors complicate the learning process, making it more demanding for algorithms to effectively learn and optimize policies in decentralized multi-agent environments.

In summary, decentralized learning and coordination in MARL present various challenges that significantly impact the convergence, stability, and efficiency of learning algorithms. The lack of global information, pronounced credit assignment problem, difficulties in maintaining policy consistency, managing the exploration-exploitation trade-off, and increased learning complexity all complicate the effectiveness of decentralized approaches. Addressing these implications requires the development of sophisticated algorithms capable of operating effectively within the constraints of decentralized settings, ensuring coherent and stable policy updates across a diverse and dynamic multi-agent system.








\section{Game Theory and MARL Integration}

In a MARL setting, the interaction of agents within a shared environment and how they influence each other's experiences and learning processes presents several challenges. Game theory provides a mathematical framework and tools to analyze these interactions, offering several insights into strategic decision-making among rational agents. This section delves into game-theoretic concepts and their integration with MARL algorithms, enhancind our understanding of complex multi-agent systems.


\subsection{Nash Equilibria in Complex Systems}
\subsubsection{Nash Equilibrium in Multi-Agent Systems}
A Nash Equilibrium represents a strategy profile $s^*$ where no agent can unilaterally improve its expected payoff by deviating from its current strategy, assuming other agents' strategies remain fixed. Consider a game with $N$ agents, each with a strategy space $S_i$ and a reward function $u_i : S_1 \times \dots \times S_N \to \mathbb{R}$. A strategy profile $s^* = (s^*_1, \dotsc, s^*_N)$ is considered a Nash Equilibrium if:

\begin{equation}
    u_i(s_i^*, s_{-i}^*) \geq u_i(s_i, s_{-i}^*), \quad \forall s_i \in S_i, \quad \forall i \in \{1, \dots, N\},
\end{equation}

\subsubsection{Integration with MARL Algorithms}

\paragraph{Nash Equilibrium in Stochastic Games}

In stochastic (Markov) games, the transition between states is based on the agents' joint actions, and the payoffs received are accumulated over time. Being each agent's objective to maximize it's expected cumulative reward $r_i$, and $S$ the set of spaces.
A Nash Equilibrium in such a context involves finding policy functions  \( \pi_i^*: S \rightarrow A_i \) where no agent can increase $R_i$ by unilaterally deviating from $\pi_i^*$, hence:

\begin{equation}
r_i( \pi_i^*, \pi_{-i}^* ) \geq r_i( \pi_i, \pi_{-i}^* ), \forall \pi_i \neq \pi_i^*
\end{equation}

This can also be written in Q-function terms, which means that for a Q-learning algorithm:

\begin{equation}
Q_i(s,  \pi_i^*, \pi_{-i}^*) = \max_{\pi_i}Q_i(s, \pi_i, \pi_{-i}^*), \forall i
\end{equation}

\paragraph{Minimax-DQN}
\cite{fan2020theoreticalanalysisdeepqlearning} In a two-player zero-sum game, agents must consider the actions of the opponent, since both are competing against each other. Borrowing the core idea behind the minimax approach, which is that each player assumes that the opponent is playing optimally, the maximizing(minimizing) player selects actions to maximize(minimize) the worst(best) possible outcome. Considering agents $i$, $j$, $u_i = -u_j$.

This assumption leads the algorithm to find the \textbf{ optimal equilibrium policies} that consider the adversarial nature of these interactions. Given a policy profile ($\pi_i^*$, $\pi_j^*$), it is \textbf{maximin}/\textbf{minimax} profile if:

\begin{equation}
Q_i(s, \pi_i^*, \pi_j^*) = \max_{\pi_i} \min_{\pi_j} Q_i(s, \pi_i, \pi_j) = \min_{\pi_j} \max_{\pi_i} Q_i(s, \pi_i, \pi_j) = - Q_j(s, \pi_i^*, \pi_j^*), \forall \pi_i, \pi_j \neq \pi_i^*, \pi_j^* 
\end{equation}

More generally, such profile is also a Nash Equilibrium. This concept can be extended to $N$ agents, where $\forall i = 1, 2, \dotsc, N$:

\begin{equation}
Q_i(s, \pi_i^*, \pi_{-i}^*) = \max_{\pi_i} \min_{\pi_{-i}} Q_i(s, \pi_i, \pi_{-i})
\end{equation}

\subsection{Evolutionary Game Theory in MARL}

The traditional approach to MARL algorithms evades the complex dynamics of systems. Evolutionary Game Theory (EGT) empathizes on the temporal evolution of strategies, which allows for the modelling of adaption and learning in Multi-Agent Reinforcement Learning. Integrating EGT with MARL provides a robust framework for understanding how agents can adapt their policies over time to achieve stable, efficient, and robust behaviors in complex, competitive, or cooperative settings.

\subsubsection{Evolutionary Dynamics in Multi-Agent Systems}

\paragraph{The Replicator Dynamics}
The Replicator Dynamics (RD) \cite{cressman2003evolutionary} are one of the core concepts of EGT. It models how the \textbf{proportion} of agents employing a particular strategy changes over time based on the fitness of the strategy relative to the population, highlighting the strategy selection process. These behaviors are modeled with \textbf{differential equations}.

For a population of agents with $m$ possible strategies, let $x_i(t)$ denote the prevalence (proportion) of the population that uses strategy $s_i$ at time $t$ and state $x(t)$. The general form of the replicator dynamic equation is given by:

\begin{equation}
\dot{x}_i(t) = x_i(t)[f_i(x(t)) - \bar{f}(x(t))]
\end{equation}

Where $f_i(x(t))$ is the fitness of a strategy $i$ at time $t$. Defined as the expected payoff when adopting strategy $i$ against the current population state $x(t)$. $\bar{f}(x(t)) = \sum_{j=1}^mx_j(t)f_j(x(t))$ is the average fitness of the population.

\paragraph{Fitness and Payoff}

In the context of EGT, fitness is equated with \textbf{payoff}, representing how successful a strategy is within the population. The payoff matrix $\mathbb{R}$ defines the expected payoff $R_{i,j}$ when strategy $i$ interacts with strategy $j$:

\begin{equation}
f_i(x(t)) = \sum_{j=1}^mA_{ij}x_j(t)
\end{equation}

This formulation models the interactions where the success of a strategy depends on its prevalence and performance against other strategies.

\subsubsection{Integration of EGT with MARL Algorithms}

\paragraph{Q-Learning on Stochastic Dispersion Games}

Several works have shown promising results on the integration of the EGT principles with MARL Algorithms.\cite{inproceedings} presented results on one-stage games, and \cite{khadka2020evolutionary} extended this work to multi-state one-stage games, i.e. stochastic Dispersion Games.
In this integration, the Q-values are interpreted as Boltzmann probabilities for the action selection. Where the \textbf{Boltzmann distribution} is the probability that a system will be in a certain state.

In a 2 players setting, each agent has a probability vector $x_1, \dotsc, x_n$ representing the likelyhood of each action over his action set $a_{1,1}, \dotsc, a_{1,n}$ and $y_1, \dotsc, y_n$ over $a_{2,1}, \dotsc, a_{2,n}$, for players $1$ and $2$, respectively. The Boltzmann distribution  for player $m$ can be formally described as:

\begin{equation}
x_{m,i}(k) = \frac{e^{\tau Q_{a_{m,i}}(k)}}{\sum_{j=1}^ne^{\tau Q_{a_{m,j}}(k)}}
\end{equation}

Where $x_{m,i}(k)$ is the probability of choosing strategy $i$ at time step $k$, and $\tau$ is the temperature that determines the \textbf{trade-off between exploration and exploitation}.

As established in \cite{10.1007/978-3-540-30115-8_18} the mutation mechanism for Q-learning can be rewritten as:

\begin{equation}
x_i\alpha\sum_jx_jln(x_j) - ln(x_i)
\end{equation}

\paragraph{Multiagent Evolutionary Reinforcement Learning (MERL)}

One notable example of this integration is the Multiagent Evolutionary Reinforcement Learning (MERL) algorithm \cite{khadka2020evolutionary}, which combines evolutionary algorithms with policy gradient methods to optimize both team-based and agent-specific objectives. Here is an overview of this algorithm.

The MERL algorithm leverages the strengths of both evolutionary strategies and reinforcement learning by maintaining two parallel optimization processes:
\begin{itemize}
    \item \textbf{Evolutionary Algorithm (EA)}: A gradient-free optimizer that focuses on maximizing the sparse team-based reward by evolving a population of team policies over generations.

    \item \textbf{Policy Gradient Method}: A gradient-based optimizer that trains individual agent policies to maximize dense, agent-specific rewards.
\end{itemize}

These two processes share information through mechanisms such as shared replay buffers and periodic migration of policies from the policy gradient optimizer to the evolutionary population.

\textbf{1- Initialization:}

\begin{itemize}
    \item \textbf{Population of Team Policies:} Initialize a population of $M$ team policies $\{ \Pi^k \}_{k=1}^M$, where each team policy $\Pi^k$ consists of $N$ agents, each with their own policy $\pi_i^k$ parameterized by $\theta_i^k$.
    \item \textbf{Shared Critic Network:} Initialize a shared critic $Q$ with parameters $\theta_Q$ to estimate the value function for policy gradient updates.
    \item \textbf{Replay Buffers:} Initialize replay buffers $\{ R_i \}_{i=1}^N$ for each agent to store experiences.
\end{itemize}

\textbf{2- Evolutionary Optimization (Team-Level Optimization)}.  At each generation if firstly performs \textbf{Rollouts and Fitness Evaluation} for each team policy $\Pi^k$ in the population:
\begin{itemize}
    \item Perform rollouts in the environment using the team policy $\Pi^k$.
    \item Collect experiences $(s_t^i, a_t^i, r_t^{i,\text{local}}, s_{t+1}^i)$ for each agent $i$ and store them in their respective replay buffers $R_i$.
    \item Accumulate the team reward $r_t^{\text{team}}$ over the episode to compute the total team fitness $f^k = \sum_{t} r_t^{\text{team}}$.
\end{itemize}

Then, \textbf{Selection and Reproduction} for each team policy:
\begin{itemize}
    \item \textbf{Selection:} Rank the team policies based on their fitness scores $f^k$. Select the top-performing policies (elites) to carry over to the next generation without modification.
    \item \textbf{Crossover and Mutation:}
    \begin{itemize}
        \item For the rest of the population, select parent policies probabilistically based on their fitness.
        \item Apply crossover by combining parts of parent policies to create offspring.
        \item Apply mutation by adding random perturbations to the offspring policies' parameters.
    \end{itemize}
\end{itemize}

\textbf{3- Policy Gradient Optimization (Agent-Level Optimization)}. Concurrently, for each agent $i$ the algorithm does \textbf{Experience Sampling}. Where it sample a minibatch of experiences $\{(s_t^i, a_t^i, r_t^{i,\text{local}}, s_{t+1}^i)\}$ from the replay buffer $R_i$.

Then, performs \textbf{Critic Update} for each agent:
\begin{itemize}
    \item Compute target values:
    \begin{equation}
    y_t = r_t^{i,\text{local}} + \gamma Q(s_{t+1}^i, \pi_i(s_{t+1}^i; \theta_i^-); \theta_Q^-),
    \end{equation}
    where $\theta_i^-$ and $\theta_Q^-$ are the parameters of the target networks.
    \item Update the critic network $Q$ by minimizing the loss:
    \begin{equation}
    L(\theta_Q) = \frac{1}{T} \sum_{t=1}^T \left( y_t - Q(s_t^i, a_t^i; \theta_Q) \right)^2.
    \end{equation}
\end{itemize}

Then, \textbf{Update the Policy:}

Update the agent's policy $\pi_i$ using the deterministic policy gradient:
\begin{equation}
\nabla_{\theta_i} J_i = \frac{1}{T} \sum_{t=1}^T \nabla_{\theta_i} \pi_i(s_t^i; \theta_i) \nabla_{a_i} Q(s_t^i, a_i; \theta_Q) \bigg|_{a_i = \pi_i(s_t^i; \theta_i)}.
\end{equation}

And finally \textbf{Update the Target Network:}

Soft update the target networks:
\begin{equation}
\theta_i^- \leftarrow \tau \theta_i + (1 - \tau) \theta_i^-,
\end{equation}
\begin{equation}
\theta_Q^- \leftarrow \tau \theta_Q + (1 - \tau) \theta_Q^-,
\end{equation}
where $\tau$ is the update rate.\newline

The evolutionary optimization in MERL reflects the Replicator Dynamics by adjusting the prevalence of team policies based on their fitness. The selection and reproduction steps ensure that strategies that yield higher team rewards become more common over time, similar to how advantageous traits spread in biological populations.

\subsection{Correlated Equilibrium and Learning in Stochastic Games}


A Correlated Equilibrium (CE) is a concept in game theory that extends the notion of Nash Equilibrium by allowing agents to coordinate their strategies through signals from an external source. This coordination can lead to higher payoffs for all agents compared to what they could achieve by acting independently. In a CE setting, agents chose their actions based on a common signal while ensuring that they have no incentive to unilaterally deviate from the recommended strategy.

\subsubsection{Mathematical formulation of correlated equilibrium}

Consider an $N$-player normal-form game. Let $A_i$ be the finite set of actions available to player $i$, and let $A = A_1 \times A_2 \times \dots \times A_N$ denote the set of all possible action profiles. Let $u_i: A \to \mathbb{R}$ be the utility function for player $i$.

A Correlated Equilibrium \cite{narahari2012correlated} is defined by a probability distribution $\lambda$ over the set of action profiles $A$ such that, for every player $i$ and for all $a_i, a_i' \in A_i$, where $a_i'$ represents any alternative action that the player could chose instead of $a_i$:

\begin{equation}
\sum_{a_{-i} \in A_{-i}} \lambda(a_i, a_{-i}) \left[ u_i(a_i, a_{-i}) - u_i(a_i', a_{-i}) \right] \geq 0,
\end{equation}

where:

\begin{itemize}
    \item $a_{-i}$ denotes the actions of all players except player $i$,
    \item $\lambda(a_i, a_{-i})$ is the probability of the action profile $(a_i, a_{-i})$.
\end{itemize}

This inequality ensures that, given the probability distribution $\lambda$, player $i$ has no incentive to unilaterally deviate from the recommended action $a_i$ to any other action $a_i'$, assuming that the other players follow the distribution $\lambda$.

\textbf{Interpretation}: The distribution $\lambda$ can be thought of as a correlation device that assigns a probability to each action profile. Before the game begins, the device sends a private recommendation to each player, suggesting an action to take. The above condition guarantees that following the recommendation is a best response for each player.

A \textbf{Nash Equilibrium} can be thought of the special case where the device recommends independent strategies. Thus, $NE \subseteq CE$ and CE potentially allow for more efficient outcomes due to coordination.

\subsubsection{Incorporating Correlated Equilibrium into MARL}

To learn Correlated Equilibria in MARL, agents can employ learning algorithms that adjust their strategies based on observed outcomes and possibly shared signals. One effective approach is using Regret Minimization algorithms, which guide agents toward strategies where they minimize their regret for not having played better in the past.

\paragraph{Regret Minimization Algorithm}
\cite{ghai2022regret} Given a policy $\pi$ obligated to approach to the optimal strategy and actions $a_k$ sampled from this policy, let $g_t^k = u_i*(a_i, a_{-i}^k) - u_i(a_i^k, a_{-i}^k)$ denote the \textbf{regret} of adopting a strategy $\pi$ in state $s$. The goal is to minimize the \textbf{cumulative regret} 
\begin{equation}
R_t^i(a_i) = \sum_{k=1}^{t-1} [ u_i(a_i, a_{-i}^k) - u_i(a_i^k, a_{-i}^k) ]
\end{equation}

The following work \cite{erez2023regret} describes an algorithm which makes use of the \textbf{value} and \textbf{Q} functions, $V^{i, \pi}(s), Q^{i, \pi}(s, a)$ to perform \textbf{Policy Optimization by Swap Regret Minimization}.

The purposed policy optimization algorithm ensures that if all agents adopt it, the state converges to a Correlated Equilibrium.

%

\paragraph{Correlated Q-Learning}

Another effective way of using Correlated Equilibrium in MARL are the family of Correlated Q-Learning algorithms \cite{greenwald2003correlated}. They extend traditional Q-learnig to enable agents to learn correlated equilibria in stochastic games.

\begin{itemize}
    \item Each agent $i$ maintains a Q-function $Q_i(s, a_1, \dots, a_N)$ representing the expected cumulative reward starting from state $s$ when agents take actions $a_1, \dots, a_N$.
    \item At each state $s$, agents compute a correlated equilibrium distribution $\lambda_s(a)$ over joint actions $a$ by solving the following linear program (LP):

    \textbf{Objective}: Maximize the expected sum of Q-values:

    \begin{equation}
    \max_{\lambda_s} \sum_{a \in A} \lambda_s(a) \left( \sum_{i=1}^N Q_i(s, a) \right)
    \end{equation}

    \textbf{Subject to}:

    \begin{itemize}
        \item \textbf{Probability Constraints}:
        \begin{equation}
        \sum_{a \in A} \lambda_s(a) = 1, \quad \lambda_s(a) \geq 0, \quad \forall a \in A
        \end{equation}
        
        \item \textbf{Correlated Equilibrium Constraints}:

        For each agent $i$ and all $a_i, a_i' \in A_i$:

        \begin{equation}
        \sum_{a_{-i} \in A_{-i}} \lambda_s(a_i, a_{-i}) \left[ Q_i(s, a_i, a_{-i}) - Q_i(s, a_i', a_{-i}) \right] \geq 0
        \end{equation}
    \end{itemize}

    \item Agents select their actions according to the computed distribution $\lambda_s(a)$.
    
    \item After observing the rewards and the next state, agents update their Q-functions:
    \[
    Q_i(s_t, a_t) \leftarrow Q_i(s_t, a_t) + \alpha \left[ r_i(s_t, a_t) + \gamma V_i(s_{t+1}) - Q_i(s_t, a_t) \right],
    \]
    where:
    \begin{itemize}
        \item $\alpha$ is the learning rate
        \item $\gamma$ is the discount factor
        \item $V_i(s_{t+1}) = \sum_{a' \in A} \lambda_{s_{t+1}}(a') Q_i(s_{t+1}, a')$
        \item $r_i$ is the reward
    \end{itemize}
\end{itemize}

The work \cite{greenwald2003correlated} extends this algorithm by introducing various different objectives, for example:
\begin{itemize}
    \item Maximize the \textit{sum} of player's rewards: $\max_{\lambda_s} \sum_{i=1}^N\sum_{a \in A} \lambda_s(a) Q_i(s, a))$
    \item Maximize the \textit{minimum} of the player's rewards: $\max_{\lambda_s} \min_{i\in N} \sum_{a \in A} \lambda_s(a) Q_i(s, a))$
    \item Maximize the \textit{maximum} of the player's rewards: $\max_{\lambda_s} \max_{i\in N} \sum_{a \in A} \lambda_s(a) Q_i(s, a))$
\end{itemize}


\subsection{Adversarial MARL and Competitive Systems}


Adversarial Multi-Agent Reinforcement Learning (Adversarial MARL) focuses on environments where agents interact with opposing entities, each striving to maximize their own objectives, often at the expense of others. Unlike cooperative settings, competitive environments require agents to anticipate and counteract the strategies of adversaries, leading to complex strategic behaviors. These interactions are prevalent in various domains such as autonomous driving, cybersecurity, and competitive games, where agents must continuously adapt to the evolving tactics of opponents to achieve optimal performance.

\subsubsection{Adversarial Dynamics in Multi-Agent Systems}

In competitive multi-agent systems, the interactions between agents can be formally modeled using game-theoretic frameworks \cite{ISHII2022252}. Consider a two-player stochastic game defined by the tuple \( \mathcal{G} = (S, A_1, A_2, P, r_1, r_2, \gamma) \), where:
\begin{itemize}
    \item \( S \) is a finite set of states.
    \item \( A_i \) is the finite set of actions available to agent \( i \) for \( i \in \{1, 2\} \).
    \item \( P: S \times A_1 \times A_2 \times S \to [0,1] \) is the state transition probability function, where \( P(s' \mid s, a_1, a_2) \) denotes the probability of transitioning to state \( s' \) from state \( s \) when agents take actions \( a_1 \) and \( a_2 \), respectively.
    \item \( r_i: S \times A_1 \times A_2 \to \mathbb{R} \) is the reward function for agent \( i \).
    \item \( \gamma \in (0,1) \) is the discount factor.
\end{itemize}

Each agent seeks to maximize their expected cumulative discounted reward:
\begin{equation}
J_i(\pi_1, \pi_2) = \mathbb{E} \left[ \sum_{t=0}^\infty \gamma^t r_i(s_t, a_t^1, a_t^2) \right],
\end{equation}
where \( \pi_i: S \to \Delta(A_i) \) is the policy of agent \( i \), mapping states to probability distributions over actions.

In adversarial settings, the reward functions are often antagonistic. For instance, in zero-sum games, the rewards satisfy:
\begin{equation}
r_1(s, a_1, a_2) = -r_2(s, a_1, a_2), \quad \forall s \in S, \ a_1 \in A_1, \ a_2 \in A_2.
\end{equation}
This ensures that the gain of one agent directly translates to the loss of the other, encapsulating the competitive nature of the interaction.

The optimal strategies in such games are derived from equilibrium concepts that account for the interdependent decision-making processes of the agents. These strategies must consider not only the immediate rewards but also the long-term consequences of actions in response to the opponent's potential strategies.

\subsubsection{Integration into MARL}

Integrating adversarial game theory concepts into Multi-Agent Reinforcement Learning involves adapting traditional RL algorithms to account for the strategic interplay between competing agents. This integration ensures that agents can learn policies that are robust against the evolving strategies of adversaries. Mathematically, this involves formulating the learning problem as a game where each agent's objective function is interdependent.

Consider the policy optimization for agent \( i \) in a competitive setting:
\begin{equation}
\theta_i^* = \arg\max_{\theta_i} \mathbb{E} \left[ \sum_{t=0}^\infty \gamma^t r_i(s_t, a_t^1, a_t^2) \right],
\end{equation}
subject to the policies of other agents \( \theta_{-i} \). The presence of adversaries introduces non-stationarity, as the environment dynamics are influenced by the learning and adaptation of opponents. To address this, the optimization incorporates opponent modeling, where each agent maintains an estimate of the strategies employed by adversaries:
\begin{equation}
\hat{\pi}_{-i}(a_{-i} \mid s) = \text{Estimator}(\text{History of } a_{-i} \text{ in state } s).
\end{equation}
Agents use these estimates to predict and counteract adversary actions, leading to more informed and strategic policy updates.

The Bellman equation for agent \( i \) in a competitive MARL setting is modified to incorporate the adversary's policy:
\begin{equation}
Q_i(s, a_i, a_{-i}) = r_i(s, a_i, a_{-i}) + \gamma \sum_{s'} P(s' \mid s, a_i, a_{-i}) \mathbb{E}_{a_i' \sim \pi_i'} [Q_i(s', a_i', a_{-i}')]],
\end{equation}
where \( \pi_i' \) denotes the updated policy of agent \( i \), and \( a_{-i}' \) are the actions of the adversary in the next state.

This formulation necessitates algorithms that can simultaneously learn optimal policies while adapting to the strategies of opponents, ensuring convergence to equilibrium strategies that are resilient in competitive environments.

\paragraph{Independent Deep Deterministic Policy Gradient (IDDPG)} 
\cite{10.1007/978-3-030-86380-7_51} \cite{marllib_ddpg} relates to the above topic by allowing each agent to learn its policy independently in a competitive setting. Mathematically, each agent \( i \) maintains its own actor \( \pi_i \) and critic \( Q_i \):
\begin{equation}
Q_i(s, a_i, a_{-i}; \theta_i^Q) = \mathbb{E} \left[ \sum_{t=0}^\infty \gamma^t r_i(s_t, a_t^i, a_t^{-i}) \mid s_0 = s, a_0^i = a_i, a_0^{-i} = a_{-i} \right].
\end{equation}
The policy update rule for agent \( i \) is:
\begin{equation}
\nabla_{\theta_i} J(\theta_i) = \mathbb{E} \left[ \nabla_{\theta_i} \pi_i(s) \nabla_{a_i} Q_i(s, a_i, a_{-i}; \theta_i^Q) \big|_{a_i = \pi_i(s)} \right].
\end{equation}
By learning independently, IDDPG allows each agent to adapt to the strategies of opponents without centralized coordination, making it suitable for environments where agents operate autonomously. 

\paragraph{Multi-Agent Deep Deterministic Policy Gradient (MADDPG)} \cite{lowe2017multi} \cite{lillicrap2015continuous} enhances IDDPG by incorporating a centralized critic that has access to the actions of all agents during training. For agent \( i \), the centralized critic \( Q_i \) is defined as:
\begin{equation}
Q_i(s, a_1, a_2; \theta_i^Q) = \mathbb{E} \left[ \sum_{t=0}^\infty \gamma^t r_i(s_t, a_t^1, a_t^2) \mid s_0 = s, a_0^1 = a_1, a_0^2 = a_2 \right].
\end{equation}
The policy gradient for agent \( i \) becomes:
\begin{equation}
\nabla_{\theta_i} J(\theta_i) = \mathbb{E} \left[ \nabla_{\theta_i} \pi_i(s) \nabla_{a_i} Q_i(s, a_1, a_2; \theta_i^Q) \big|_{a_i = \pi_i(s)} \right].
\end{equation}
This centralized training approach allows the critic to better estimate the value of actions in the presence of adversaries, leading to more informed and effective policy updates.

\paragraph{Fully Actor-Critic MAC (FACMAC)} \cite{DBLP:journals/corr/abs-2003-06709}  \cite{DBLP:journals/corr/FoersterAFW16a} builds upon the MADDPG framework by employing fully actor-critic architectures with centralized training and decentralized execution. Each agent \( i \) maintains:
\begin{equation}
\pi_i(s; \theta_i^\pi) \quad \text{and} \quad Q_i(s, a_1, a_2; \theta_i^Q).
\end{equation}
The centralized critic in FACMAC leverages the full action space to provide comprehensive feedback:
\begin{equation}
Q_i(s, a_1, a_2; \theta_i^Q) = \mathbb{E} \left[ \sum_{t=0}^\infty \gamma^t r_i(s_t, a_t^1, a_t^2) \mid s_0 = s, a_0^1 = a_1, a_0^2 = a_2 \right].
\end{equation}
The policy gradient update for agent \( i \) is:
\begin{equation}
\nabla_{\theta_i^\pi} J(\theta_i^\pi) = \mathbb{E} \left[ \nabla_{\theta_i^\pi} \pi_i(s) \nabla_{a_i} Q_i(s, a_1, a_2; \theta_i^Q) \big|_{a_i = \pi_i(s)} \right].
\end{equation}
FACMAC's fully centralized critics enable agents to better navigate the complexities of competitive environments by providing detailed evaluations of joint actions, facilitating more strategic policy learning.

\paragraph{Learning with Opponent-Learning Awareness (LOLA)} \cite{DBLP:journals/corr/abs-1709-04326} is an advanced algorithm in Adversarial MARL that enables agents to consider the learning dynamics of their opponents during policy updates. Unlike traditional MARL algorithms that treat opponents as part of a non-stationary environment, LOLA explicitly models and anticipates the policy updates of adversaries, allowing agents to adjust their strategies proactively. This mutual awareness fosters more sophisticated and stable strategic behaviors, enhancing the effectiveness of agents in competitive settings.

LOLA agents incorporate higher-order gradients to account for the impact of their policy updates on the learning trajectories of opponents. For agent \( i \) interacting with agent \( j \), the policy update rule is defined as:
\begin{equation}
\theta_i^{\text{new}} = \theta_i + \alpha \nabla_{\theta_i} J_i + \beta \nabla_{\theta_i} \left( \nabla_{\theta_j} J_j \cdot \nabla_{\theta_i} J_i \right),
\end{equation}
where:
\begin{itemize}
    \item \( \alpha \) is the primary learning rate.
    \item \( \beta \) is the meta-learning rate that determines the influence of opponent learning.
    \item \( J_i \) is the objective function of agent \( i \).
    \item \( \theta_j \) are the policy parameters of agent \( j \).
\end{itemize}

The term \( \beta \nabla_{\theta_i} \left( \nabla_{\theta_j} J_j \cdot \nabla_{\theta_i} J_i \right) \) captures the anticipated effect of agent \( i \)'s policy update on agent \( j \)'s learning process. By incorporating this term, agent \( i \) can adjust its policy in a manner that not only maximizes its own objective but also considers how its actions influence the strategies of adversaries.

\paragraph{Generative Adversarial Imitation Learning (GAIL)}

\cite{DBLP:journals/corr/HoE16} Generative Adversarial Imitation Learning (GAIL) leverages adversarial training frameworks to enable agents to learn policies by imitating expert demonstrations. GAIL formulates imitation learning as a two-player game between a generator (the policy to be learned) and a discriminator that distinguishes between the policy's actions and expert actions. This adversarial setup ensures that the learned policy closely mimics the expert behavior without requiring explicit reward engineering.

GAIL's objective is to train a policy \( \pi \) that generates behavior indistinguishable from expert demonstrations \( \pi_E \). This is achieved through the following minimax optimization problem:
\begin{equation}
\min_\pi \max_D \mathbb{E}_{(s,a) \sim \pi} [ \log D(s, a) ] + \mathbb{E}_{(s,a) \sim \pi_E} [ \log (1 - D(s, a)) ] - \lambda H(\pi),
\end{equation}
where:
\begin{itemize}
    \item \( D(s, a) \) is the discriminator that aims to distinguish between actions taken by the policy \( \pi \) and the expert policy \( \pi_E \).
    \item \( H(\pi) \) is an entropy regularization term that encourages exploration.
    \item \( \lambda \) is a regularization parameter balancing the entropy term.
\end{itemize}

\textbf{Discriminator Update}:
\begin{equation}
D^* = \arg\max_D \mathbb{E}_{(s,a) \sim \pi_E} [ \log (1 - D(s, a)) ] + \mathbb{E}_{(s,a) \sim \pi} [ \log D(s, a) ].
\end{equation}
The discriminator learns to differentiate between expert and policy-generated actions by maximizing the likelihood of correctly classifying each sample.

\textbf{Policy Update}:
\begin{equation}
\theta_\pi \leftarrow \theta_\pi + \alpha \nabla_{\theta_\pi} \mathbb{E}_{(s,a) \sim \pi} [ \log D(s, a) ],
\end{equation}
where \( \alpha \) is the learning rate for the policy parameters \( \theta_\pi \). The policy aims to maximize the probability of the discriminator misclassifying its actions as expert actions, effectively imitating the expert behavior.

In competitive environments, GAIL can be utilized for agents to learn policies that mimic expert adversaries, thereby enhancing their ability to counteract sophisticated opponent strategies. By training the policy to fool the discriminator into believing its actions are from an expert, the agent develops robust and effective behaviors that can adapt to and counteract adversarial tactics.


\section{Conclusions}

This paper has presented an in-depth examination of advanced topics in Multi-Agent Reinforcement Learning (MARL), focusing on key challenges and their intersection with game theory. Our analysis demonstrates that the integration of game-theoretic concepts with MARL algorithms provides powerful frameworks for addressing fundamental challenges in multi-agent learning. Concepts such as Nash equilibria, evolutionary dynamics, and correlated equilibrium offer mathematical foundations that are essential for developing more robust and effective learning strategies in multi-agent systems.

Non-stationarity and partial observability continue to present substantial obstacles, requiring increasingly sophisticated approaches that can adapt to changing dynamics and incomplete information. Through our research, we have shown that the incorporation of game-theoretic principles helps in designing algorithms that can better handle these challenges through strategic reasoning and equilibrium-based solutions. While scalability issues in large agent populations can be mitigated through decentralized learning approaches, this solution introduces new challenges in coordination and policy consistency.

The synthesis of game theory and MARL continues to open new avenues for advancing artificial intelligence in multi-agent systems, particularly in areas requiring complex strategic interactions and decision-making under uncertainty.

\bibliographystyle{plainnat} 
\bibliography{references}

\end{document}